\documentclass{article}

\usepackage{authblk}
\usepackage[font=small]{caption}

\usepackage{float} 
\usepackage[export]{adjustbox}% by Fritz
\usepackage{amsthm} % added by Fritz
\usepackage{braket} % added by Fritz

\usepackage{placeins}

\usepackage{amsmath}
\usepackage{amsfonts}
\usepackage{amssymb}

\usepackage{booktabs}

\usepackage[separate-uncertainty=true,multi-part-units=single]{siunitx} % added by Fritz
\usepackage[final]{changes} % does not work for the part that is included from the appendix - seach for "here123"
\usepackage{bm}
\let\citeleft=(
\let\citeright=)

\begin{document}

\pdfinfo{
   /Author (AUTHORS)
   /Title (TITLE)
}

%%TC:ignore
%\title{\vspace{-2cm}TITLE}

\title{\vspace{-2cm} Unbiased Signal Equation for Quantitative Magnetization Transfer Mapping in Balanced Steady-State Free Precession MRI}
\author[1,3]{Fritz M. Bayer}
\author[2]{Michael Bock}
\author[1]{Peter Jezzard}
\author[1]{Alex K. Smith}
\affil[1]{\small Wellcome Centre for Integrated Neuroimaging, FMRIB Division, Nuffield Dept of Clinical Neurosciences, University of Oxford, Oxford, UK}
\affil[2]{\small Dept. of Radiology, Medical Physics, Medical Center – University of Freiburg, Faculty of Medicine, University of Freiburg, Freiburg, Germany}
\affil[3]{\small D-BSSE, ETH Zurich, Mattenstrasse 26, 4058 Basel, Switzerland}

%\author[1]{AUTHOR 1}
%\author[2]{AUTHOR 2}
%
%\affil[1]{\small AFFILIATION 1}
%\affil[2]{\small AFFILIATION 2}
\maketitle

\vfill
\noindent
\textit{Running head:} Unbiased Signal Equation for Quantitative Magnetization Transfer Mapping in bSSFP

\noindent
\textit{Address correspondence to:} \\
  Peter Jezzard, WIN Centre FMRIB Division, John Radcliffe Hospital, Headley Way, Headington, Oxford, OX3 9DU, UK. \\
  peter.jezzard@univ.ox.ac.uk

\noindent
This work was supported by Wellcome Trust Grant/Award Number: 203139/Z/16/Z; Dunhill Medical Trust; Oxford NIHR Biomedical Research Centre; Whitaker International Program; Cusanuswerk Scholarship.

\noindent
Approximate word count: 225 (Abstract) 3362 (body)\\

\noindent
Submitted to \textit{Magnetic Resonance in Medicine} as a Full Paper.

\clearpage

\section*{Abstract}

\noindent
\textbf{Purpose}: Quantitative magnetization transfer (qMT) imaging can be used to %\deleted[]{detect the signal of protons attached to relatively immobile macromolecules} 
\added[remark={R1.1}]{quantify the proportion of protons in a voxel attached to macromolecules}. Here, we show that the original qMT balanced steady-state free precession (bSSFP) model is biased due to over-simplistic assumptions made in its derivation. 

\noindent
\textbf{Theory and Methods}: We present an improved model for qMT bSSFP, which incorporates finite radio-frequency (RF) pulse effects as well as simultaneous exchange and relaxation. Further, a correction to finite RF pulse effects for sinc-shaped excitations is derived. The new model is compared to the original one in numerical simulations of the Bloch-McConnell equations and in previously acquired in-vivo data.

\noindent
\textbf{Results}: Our numerical simulations show that the original signal equation is significantly biased in typical brain tissue structures (by 7-\SI{20}{\percent}) whereas the new signal equation outperforms the original one with minimal bias ($< \SI{1}{\percent}$).  It is further shown that the bias of the original model strongly affects the acquired qMT parameters in human brain structures, with differences in the clinically relevant parameter of pool-size-ratio of up to \SI{31}{\percent}. Particularly high biases of the original signal equation are expected in an MS lesion within diseased brain tissue (due to a low T2/T1-ratio), demanding a more accurate model for clinical applications.

\noindent
\textbf{Conclusion}: The improved model for qMT bSSFP is recommended for accurate qMT parameter mapping in healthy and diseased brain tissue structures.
  
\noindent
\textbf{Keywords}: magnetization transfer, balanced SSFP, quantitative imaging
%%TC:endignore

\clearpage

\section*{Introduction}

\noindent
Quantitative magnetization transfer (qMT) imaging can be used to %\deleted[]{detect the signal of protons attached to relatively immobile macromolecules} 
\added[remark={R1.1}]{quantify the proportion of protons in a voxel attached to macromolecules}. qMT  has shown considerable promise for characterising myelin-related diseases, such as multiple sclerosis. Due to a high signal-to-noise ratio and short acquisition times, balanced steady state free precession (bSSFP) acquisition modules have become a popular method for quantifying MT parameters \cite{gloorQuantitativeMagnetizationTransfer2008,gloorIntrascannerInterscannerVariability2011,garciaFastHighresolutionBrain2012}. However, the derivation of the qMT bSSFP signal equation is based on two major assumptions, which limit its generality and accuracy.

Firstly, it is assumed that magnetisation relaxation and the spin exchange between the free and macromolecular pool (MT) can be modelled as independent processes. This implies that the continuous phenomenon of MT has an instantaneous effect on the magnetisation. Though the separation of exchange and relaxation simplifies the derivation of the original qMT bSSFP signal equation, this assumption does not accurately describe the physical nature of MT, as these effects happen simultaneously.

Furthermore, the originally proposed signal equation assumes an instantaneous rotation of magnetisation by the RF pulse. Bieri and Scheffler have shown \cite{bieriSSFPSignalFinite2009, bieriAnalyticalDescriptionBalanced2011} that this assumption does not accurately describe the finite nature of an RF pulse in bSSFP due to an overestimation of transverse relaxation. While this effect is negligible for short pulse durations $\frac{T_{RF}}{TR} \ll 1$, a significant bias is introduced if that condition is not satisfied \cite{bieriSSFPSignalFinite2009}. In conventional bSSFP (non-qMT), this bias can amount to \SI{10}{\percent} ($\alpha$ $\sim$ $90^{\circ}$, T2/T1 $\ll$ 1) \cite{bieriSSFPSignalFinite2009, bieriAnalyticalDescriptionBalanced2011}. As qMT bSSFP is based on a stepwise variation of the RF pulse duration, this condition is certainly not met in the original qMT bSSFP acquisition scheme, where the ratio $\frac{T_{RF}}{TR}$ can be as high as 0.44 \cite{gloorQuantitativeMagnetizationTransfer2008, gloorFiniteRFPulse2010}. A correction to this bias has been proposed for Gaussian pulse shapes, which are, however, not commonly used in qMT bSSFP, where a sinc pulse is more typically used \cite{gloorQuantitativeMagnetizationTransfer2008, garciaFastHighresolutionBrain2012, gloorIntrascannerInterscannerVariability2011}.

Here, we present an improved signal equation for qMT bSSFP, which incorporates finite pulse effects as well as simultaneous exchange and relaxation. A correction to finite RF pulse effects for sinc-shaped excitations is derived. By means of numerical simulations of the Bloch-McConnell equations, it is demonstrated that the original signal equation is significantly biased in typical brain tissue structures. Additionally, this bias is strongly dependent on the time-bandwidth (TBW) product for sinc pulses; thus, a framework to minimise this bias is presented.

%By means of numerical simulations of the Bloch-McConnell equations, it is demonstrated that the original signal equation is significantly biased in typical brain tissue structures (by 7-\SI{20}{\percent}). In contrast, the new signal equation outperforms the original one with minimal bias ($<$ \SI{1}{\percent}). Additionally, it is demonstrated that the bias is strongly dependent on the time-bandwidth product for sinc-shaped RF pulses and a framework to minimise this bias is presented.
\section*{Refined Balanced SSFP Signal Equation}\label{sec:derivation}

\noindent
% Consider two pools of magnetization $M_f=[M_{xf},M_{yf},M_{zf}]^T$ and $M_m=[M_{zm}]^T$ with longitudinal and transversal exchange rates $R_{1,j}=\frac{1}{T_{1,j}}$ and $R_{2,i}=\frac{1}{T_{2,j}}$ for $i=m,f$. 
In this section, a new qMT bSSFP signal equation is derived allowing for simultaneous magnetization exchange and relaxation, and correcting for the instantaneous rotation by the RF pulse. To model the magnetisation dynamics a single bSSFP acquisition cycle of duration $TR$ is considered, that is repeated until steady-state is reached. Each cycle can be split into two epochs:
\renewcommand{\theenumi}{\Roman{enumi}}
\begin{enumerate}
	\item Excitation by the RF pulse
	\item Free relaxation (including spin information exchange between pools)
\end{enumerate}
To derive the magnetisation at steady-state, each epoch can be modelled independently and subsequently unified by the steady-state condition.

Analogous to the original derivation \cite{gloorQuantitativeMagnetizationTransfer2008}, the excitation by the RF pulse (epoch I.) is initially assumed to be instantaneous $T_{RF}\rightarrow 0$ (correction follows below). Thus the magnetisation state is instantly rotated at $t=nTR$, $n\in \mathbb{N}_0$, which is described by the following formalism
\begin{equation*}
M(t'=nTR) = 
\begin{cases} 
M^-(n) & \text{before rotation via the RF pulse} \\
M^+(n) & \text{after rotation via the RF pulse}
\end{cases}
\end{equation*}
This convention was established by Freeman \cite{freemanPhaseIntensityAnomalies1971} and is commonly used in bSSFP. The RF pulse leads to a rotation of the free-pool magnetisation around the x-axis and can therefore be modelled via the rotation matrix $R_x(\alpha)$, representing a clockwise rotation in the x-plane with angle $\alpha$ for an anticlockwise polarized RF field. Simultaneously, the pulse saturates the macromolecular pool, which can be modelled using the mean saturation rate $\langle W(\Delta \rightarrow 0 ) \rangle$. Thus, the operator, representing the action of the pulse on the magnetisation $M=[M_{xf},M_{yf},M_{zf},M_{zm}]^T$, is given by \added[remark={R1.4}]{}
\begin{equation}
\added{\bm{R_x(\alpha, t)}}= 
\begin{bmatrix}
1 & 0 & 0 & 0 \\
0 & \cos \alpha & \sin \alpha & 0\\
0 & -\sin \alpha & \cos \alpha & 0\\
0 & 0&0&e^{-\langle W(\Delta \rightarrow 0 ) \rangle t}
\end{bmatrix} 	
\end{equation}
satisfying the relation \added[remark={R1.4}]{}
\begin{equation}\label{equ:2}
M^-(n)=\added{\bm{R_x}} M^+(n)
\end{equation}
The mean saturation rate $\langle W(\Delta)\rangle$ used in this derivation is equivalent to the one proposed in the work by Gloor and is described in detail elsewhere \cite{gloorQuantitativeMagnetizationTransfer2008}. 

During free relaxation (epoch II.), the magnetisation $M=[M_{xf},M_{yf},M_{zf},M_{zm}]^T$ can be modelled by solving the Bloch-McConnell Equations in the unperturbed case
\begin{equation}\label{equ:bme}
\frac{dM(t')}{dt'}= 
\begin{bmatrix}
-R_{2f} \!\!\! & 0 & 0 & 0\\
0 & \!\!\!-R_{2f} & \omega_1 & 0\\
0 & \!\!-\omega_1 & \!\! -\big(R_{1f}+k_{fm}\big) \!\!\!\!\! & k_{mf}\\
0 & 0 & k_{fm} & \!\!\!\!\! -\big(R_{1m}+k_{fm}\big)
\end{bmatrix}
M(t')  + 
\begin{bmatrix}
0\\
0\\
R_{1f}M_{0f}\\
R_{1m}M_{0f}F
\end{bmatrix}
\end{equation}
where ${t' = t - n \cdot TR}$ with $n \in \mathbb{N}_0$ is the time of the $n^\text{th}$ acquisition cycle and the magnetisation is $M=[M_{xf},M_{yf},M_{zf},M_{zm}]^T$. \added[remark={R2.3}]{$R_{1f}$ and $R_{2f}$ are the longitudinal and transversal relaxation rates of the free pool, respectively, $k_{mf}$ and $k_{fm}$ are the exchange rates from macromolecular to free pool and free to macromolecular pool, respectively, and the pool-size-ratio $F = M_{0m}/M_{0f}$ describes the ratio of the free pool $M_{0f}$ and the macromolecular pool $M_{0m}$.}
%$F = M_{0m}/M_{0f}$: pool-size-ratio, $k_{mf}$: transfer of longitudinal magnetization from macromolecular to free pool, $R_{1f}$: longitudinal exchange rate, $T_{2f}$: longitudinal relaxation time of the free pool.x
%\added[remark={R1.3}]{ % here123
	Within the range ${n\cdot TR<t<(n+1)\cdot TR}$, Equation \ref{equ:bme} results in a \textit{first order linear inhomogeneous matrix ODE}
	\begin{equation}
	\frac{dM(t')}{dt'}=\bm{\xi_1} M(t')+\bm{\xi_2} M_0
	\end{equation}
	as the relaxation and exchange matrix $\bm{\xi_1}(t')=\bm{\xi_1}$ is time-independent. The solution to this \textit{first order linear inhomogeneous matrix ODE} is given by
	\begin{equation}\label{equ:1}
	M(t)=e^{\bm{\xi_1}t}M(t=0)+\bm{\xi_1}^{-1}(e^{\bm{\xi_1}t}-\bm{I})\bm{\xi_2} M_0
	\end{equation}
	%where the proof is attached in appendix \ref{ch:proofint}.
	%
	For repeated iterations of the pulse ($n \rightarrow \infty$), the magnetisation reaches a dynamic steady-state, satisfying the condition
	\begin{equation}\label{equ:ss}
	M^-(n+1)=\bm{R_z}M^-(n) \Leftrightarrow M^+(n+1)=\bm{R_z}M^+(n)
	\end{equation}
	where the rotation matrix $\bm{R_z}:=\bm{R_z}(\Phi=180^{\circ})$ represents the change in sign of the flip angle after each iteration
	\begin{equation}
	\bm{R_z}(\Phi=180^{\circ})=
	\begin{bmatrix}
	-1&\;0\;&\;\; 0\;&\;\; 0\;\; \\
	\;0&-1\;&\;\; 0\;&\;\; 0\;\; \\
	\;0&\;0\;&\;\;1\;&\;\; 0\;\;\\
	\;0&\;0\;&\;\;0\;&\; \; 1\;\;
	\end{bmatrix}	
	\end{equation}
	The magnetisation during one pulse cycle (epochs I. and II.), can be modelled by combining Equations \ref{equ:2} and \ref{equ:1}. This allows to relate the magnetisation before the $(n+1)^{\text{th}}$ pulse to the magnetization before the $n^{\text{th}}$ pulse
	\begin{equation}
	M(n+1)^-=e^{\bm{\xi_1} TR-\bm{I}} \bm{R_x} M(n)^-+\bm{\xi_1}^{-1}(e^{\bm{\xi_1} TR}-\bm{I})\bm{\xi_2} M_0
	\end{equation}
	This equation can be solved under the dynamic steady-state condition (Equation \ref{equ:ss}) for $n\rightarrow \infty$ with the Ansatz
	\begin{equation}
	\bm{R_z}M(n)^-=M(n+1)^-
	\end{equation}
	\begin{equation}
	\Leftrightarrow \qquad \bm{R_z} M(n)^-=e^{\bm{\xi_1} TR-\bm{I}} \bm{R_x} M(n)^-+\bm{\xi_1}^{-1}(e^{\bm{\xi_1} TR}-\bm{I})\bm{\xi_2} M_0
	\end{equation}
	\begin{equation}
	\Leftrightarrow \qquad M(n)^-=(\bm{R_z}-e^{\bm{\xi_1} TR}\bm{R_x})^{-1} \bm{\xi_1}^{-1}(e^{\bm{\xi_1} TR}-\bm{I})\bm{\xi_2} M_0
	\end{equation}
	resulting in the solution at steady-state
	\begin{equation}
	M_{SS}=M(n\rightarrow \infty)^+=\bm{R_x}(\bm{R_z}-e^{\bm{\xi_1} TR}\bm{R_x})^{-1} \bm{\xi_1}^{-1}(e^{\bm{\xi_1} TR}-\bm{I})\bm{\xi_2} M_0
	\end{equation}
%} % here123

%\deleted[]{As proven in the Appendix \ref{sec:A1}, the solution to this equation at steady-state is given by
%\begin{equation}
%M_{SS}=M(n\rightarrow \infty)^+=R_x(R_z-e^{\xi_1 TR}R_x)^{-1} \xi_1^{-1}(e^{\xi_1 TR}-I)\xi_2 M_0
%\end{equation}
%}

The operator \added[remark={R1.4}]{}$\added{\bm{R_x}}(\alpha, t)$ represents an instant rotation of the magnetisation in the free pool. This assumption is commonly used in MRI, but leads to an overestimation of transverse relaxation during excitation \cite{bieriSSFPSignalFinite2009,bieriAnalyticalDescriptionBalanced2011}. Throughout the rotation caused by a pulse of finite pulse duration, the magnetisation spends a period when it has parallel alignment with the static magnetic field, i.e. its equilibrium orientation. This reduces the transverse relaxation, which can be accounted for by the correction suggested by Bieri for the one-pool model \cite{bieriSSFPSignalFinite2009}
\begin{equation}\label{equ:cor}
R_2 \rightarrow \tilde{R_2}=\bigg(1-\zeta \frac{T_{RFE}}{TR}\bigg)R_2, \; \; \forall \: T_{RFE}>0
\end{equation}
with
\begin{equation}
\zeta \approx 0.68-0.125 \bigg(1+ \frac{T_{RFE}}{TR}  \bigg)  \frac{R_1}{R_2}
\end{equation}
where the hard pulse time equivalent $T_{RFE}$ is a pulse-shape-dependent constant. While this constant has previously been derived for Gaussian pulse shapes, here we present a solution for sinc pulse shapes (details in Appendix), as these are commonly used in qMT bSSFP
\begin{equation}\label{equ:Trefall}
T_{RFE} = 
\begin{cases} 
T_{RF} & \text{for hard pulse (by definition)} \\
1.20 \cdot \frac{T_{RF}}{TBW} & \text{for Gaussian pulse (proof in \cite{bieriSSFPSignalFinite2009})}\\
\frac{4 T_{RF}}{TBW \pi} \frac{1- \cos \big(\frac{\pi TBW}{2} \big)}{\text{Si}\big( \frac{\pi TBW}{2} \big)} & \text{for sinc pulse (proof in Appendix}
\end{cases}
\end{equation}
Here, Si denotes the sine integral defined in Equation \ref{equ:sineint}. The correction accounts for the overestimation of transverse relaxation during excitation and therefore considers the finite pulse duration; the derivation to the correction given by Equation \ref{equ:cor} can be found at \cite{bieriSSFPSignalFinite2009}.

As the finite pulse duration correction only affects the transverse magnetisation, which is generally neglected in the macromolecular pool, the two-pool model can be corrected by transforming $R_2$ within the matrix \added[remark={T1.4}]{}$\added{\bm{\tilde{\xi_1}}}=\added{\bm{\xi_1}}(R_{2f}\rightarrow \tilde{R}_{2f})$
\begin{equation}
\added{\bm{\tilde{\xi}_1}}=
\begin{bmatrix} 
-\big(1-\zeta \frac{T_{RFE}}{TR}\big)R_{2f} \!\! & 0 & 0 & 0 \\
0 & \!\!\! -\big(1-\zeta \frac{T_{RFE}}{TR}\big)R_{2f} \!\!\! & \omega_1 & 0 \\ 
0 & -\omega_1 & \!\!\! -\big(R_{1f}+k_{fm}\big) \!\!\! & k_{mf}\\
0 & 0 & k_{fm} & \!\! -\big(R_{1m}+k_{fm}\big)
\end{bmatrix}
\end{equation}
As $M_{xf}$ is decoupled from the other components, the coupled equations can be reduced to $M=[M_{yf}, M_{zf},M_{zm}]^T$.
This leads to the corrected steady-state solution for bSSFP in matrix notation, taking into account finite pulse duration effects and concurrent magnetization exchange and relaxation\added[remark={T1.4}]{}
\begin{equation}\label{equ:Mss}
M_{SS}=\tilde{R}_x(R_z-e^{\tilde{\xi}_1 TR}\tilde{R}_x)^{-1} \tilde{\xi}_1^{-1}(e^{\tilde{\xi}_1 TR}-\added{\bm{I}})\xi_2 M_0
\end{equation}
%\qed
%The \textit{refined signal equation} \ref{equ:Mss} allows the magnetisation during bSSFP acquisition to be modelled in presence of magnetisation transfer and gives an alternative to the \textit{original signal equation} \ref{equ:gloor}.
Note that while Equation \ref{equ:Mss} describes the magnetisation as a whole, experimentally only the transverse component of the free pool $M_{yf}$ is measured.

\subsection*{Numerical Studies}\label{sec:sim}

\noindent
To validate the analytical solution, simulations were performed by numerically solving the Bloch-McConnell Equations for typical brain tissue parameters (Table \ref{tab:1}). % The simulation has been split into intervals of excitation by the RF pulse and free relaxation. 
% As illustrated in Figure \ref{fig:sim}, the iteration of these intervals converges to a constant steady-state magnetisation. An initial oscillation of $|M_{xy}|$ can be observed, which is a consequence of the change in sign of the flip angle ($+\alpha,-\alpha,+\alpha,...$) after each repetition. 

Similar to the originally proposed acquisition, the flip angle $\alpha$ and the pulse duration $T_{RF}$ have been varied while setting all other acquisition parameters constant. % As shown in Figure \ref{fig:sim}, this results in different steady-state magnetisations.
%\begin{figure}[t]
%	\centering
%	\input{figures/plot4.tex}
%	\caption[Simulation of Magnetisation at %Steady-State]{Illustrative simulation of transversal magnetisation $M_{xy}$ during the first 80 pulse iterations for varied flip angles $\alpha$ and constant pulse duration $T_{RF}$ (left) and vice versa (right). Constant acquisition parameters are $T_{RF}=\SI{0.3}{ms}$ and $\alpha=\SI{35}{°}$.}
%	\label{fig:sim}
%\end{figure}
As suggested in the original paper by Gloor \cite{gloorQuantitativeMagnetizationTransfer2008}, an on-resonance ($\Delta \omega = 0$) sinc pulse shape has been chosen for excitation (Equation \ref{equ:sinc}).

The effect of the sinc pulse on the free pool magnetisation, given by $\omega_1$, has been simulated based on
\begin{equation}
\omega_1= \gamma B_1 =\begin{cases} 
\gamma A(\alpha, t_0) \text{sinc}\Big(\frac{\pi t}{t_0}\Big)  &\;\; \text{for} \;\; n TR-T_{RF} < t < nTR+T_{RF} \\
0 &\;\; \text{elsewhere}
\end{cases}
\end{equation}
where the amplitude of each pulse has been calculated according to
\begin{equation}
A(\alpha,t_0)=2 \pi \alpha \bigg(360^{\circ} \gamma \int_{-T_{RF}/2}^{T_{RF}/2} \text{sinc}\big(\frac{\pi t}{t_0}\big) dt \bigg)^{-1}
\end{equation}
and the half-width of the central lobe $t_0$ is related to $T_{RF}$ and $TBW$ according to Equation \ref{equ:TBW}. Due to its superior performance in tissue \cite{morrisonModelingMagnetizationTransfer1995}, a super-Lorentzian lineshape has been chosen for absorption according to
\begin{equation}
R_{RF}=\begin{cases} 
\pi \gamma^2B_1^2g_m(\Delta \omega,T_{2m})  &\;\; \text{for} \;\; n TR-T_{RF} < t < nTR+T_{RF} \\
0 &\;\; \text{elsewhere}
\end{cases}
\end{equation}
for the $n^\text{th}$ iteration $n\in \{1,2,...,N\}$ and
\begin{equation}
g_m(\Delta \omega,T_{2m})=\sqrt{\frac{\pi}{2}}\int_{0}^{\frac{\pi}{2}}\frac{T_{2m}}{|3u^2-1|}\exp\Bigg(-2\bigg(\frac{2\pi \Delta \omega T_{2m}}{3u^2-1}\bigg)^2\Bigg)du
\end{equation}
Note that the super-Lorentzian absorption lineshape has a singularity at $g_m(\Delta \omega=0)$. Analogous to previous studies \cite{bieriOriginApparentLow2006, gloorQuantitativeMagnetizationTransfer2008}, the absorption lineshape has been extrapolated from \SI{1}{kHz} to the asymptotic limit, resulting in $g_m(\Delta \omega \rightarrow 0) = \SI{1.4e-5}{s}$ for which a constant $T_{2m}=\SI{12}{\micro s}$ has been assumed.

\subsection*{In-Vivo Studies}

\noindent
In addition to the simulations, the performance of the refined signal equation in comparison to the original one of Gloor et al. \cite{gloorQuantitativeMagnetizationTransfer2008} was investigated in previously acquired human brain data. The images used in this work have been taken from an open source publication by Cabana et al. \cite{cabanaQuantitativeMagnetizationTransfer2015}. They were acquired for a single volunteer at 1.5 T (Siemens Healthineers, Erlangen, Germany) using the standard protocol of varied flip angles and pulse durations, suggested in the original publication on qMT bSSFP \cite{gloorQuantitativeMagnetizationTransfer2008}. The bSSFP acquisition parameters were varied as follows:
\renewcommand{\theenumi}{\roman{enumi}}
\begin{enumerate}
	\item Eight bSSFP sequences with constant flip angle $\alpha= 35^{\circ}$ and varied pulse duration $T_{RF} =\SI{0.23}{ms}$, \SI{0.3}{ms}, \SI{0.4}{ms}, \SI{0.58}{ms}, \SI{0.84}{ms}, \SI{1.2}{ms}, \SI{1.6}{ms}, \SI{2.1}{ms}. 
	\item Eight bSSFP sequences with constant pulse duration $T_{RF}=\SI{0.27}{ms}$ and varied flip angle $\alpha=5^{\circ}$, $10^{\circ}$, $15^{\circ}$, $20^{\circ}$, $25^{\circ}$, $30^{\circ}$, $35^{\circ}$, $40^{\circ}$. 
\end{enumerate}
The repetition time $TR$ of each single sequence was chosen such that $T_d=TR-T_{RF}=\SI{2.7}{ms}$ remained constant and a sinc-shaped RF pulse of $TBW=2.7$ was used for excitation. A field-of-view (FOV) of $\SI{256}{mm} \times \SI{256}{mm} \times \SI{32}{mm}$ with acquisition matrix $128 \times 128 \times 16$ was selected. Additionally, $T_1$ maps were acquired using two SPGR sequences \added[remark={R2.1}]{with $TR/TE = \SI{9.8}{ms}/\SI{4.77}{ms}$, bandwidth = \SI{140}{Hz/Pixel} and varying flip angles of $\alpha=4^{\circ}$ and $\alpha=15^{\circ}$} according to the DESPOT1 method \cite{gloorQuantitativeMagnetizationTransfer2008, deoniHighresolutionT1T22005}.

% (TR/TE = 9.8 ms/4.77 ms, bandwidth = 140 Hz/Pixel) with flip angles α = 4° and α = 15°

%0.23, 0.3, 0.4, 0.58, 0.84, 1.2, 1.6, 2.1 ms
%2.92, 2.99, 3.09, 3.26, 3.53, 3.88, 4.28, 4.78 ms
%5, 10, 15, 20, 25, 30, 35, 40°

\subsection*{Quantitative MT Parameter Analyis}\label{sec:analysis}

%The refined, as well as the original, qMT bSSFP signal equation are dependant on eight parameters, from which three are fixed by the acquisition ($TR$, $T_{RF}$ and $\alpha$). $T_{1f}$ is calculated from the additional SPGR acquisition by DESPOT1 and due to the insensitivity of the model on $T_{1m}$ \cite{gloorQuantitativeMagnetizationTransfer2008}, $T_{1m}$ is set equal to $T_{1f}$. $F$, $k_{mf}$ and $T_{2f}$ fitted.

\noindent
In order to determine the qMT parameters in each voxel, the qMT bSSFP signal equation was fitted to the acquired steady-state magnetisation by means of a non-linear least-squares fit. Both refined and original qMT bSSFP signal equations are dependent on five parameters: $F$, $k_{mf}$, $T_{1f}$, $T_{2f}$ and $T_{1m}$. However, the additional acquisition of $T_{1f}$ allows that parameter to be fixed and $T_{1m}$ can be set equal to $T_{1f}$ due to its insensitivity to the magnetisation \cite{gloorQuantitativeMagnetizationTransfer2008, henkelmanQuantitativeInterpretationMagnetization1993, yarnykh2012fast}\added[remark={R2.2}]{}. Thus, the remaining parameters $F$, $k_{mf}$ and $T_{2f}$ were fitted on a voxel-by-voxel basis \added[remark={R2.5}]{within the ranges $0.01 \le F \le \SI{30}{\percent}$, $0.0001 \le k_{mf} \le \SI{100}{s^{-1}}$ and $0.01 \le T_{2f} \le \SI{0.2}{s}$ using the starting points $\hat{F}=\SI{10}{\percent}$,  $\hat{k}_{mf}=\SI{30}{s^{-1}}$ and $\hat{T}_{2f}=\SI{0.04}{s}$}. \added[remark={R1.6}]{$M_{0f}$ has been set to one as has been done previously \cite{smith2014rapid}.} All computations were performed in Matlab (MathWorks, Natick, MA) and code has been partially taken from the qMRlab toolbox \cite{cabanaQuantitativeMagnetizationTransfer2015}. A non-parametric Wilcoxon signed-rank test was used for statistical testing. 

%\added[remark={R2.5}]{F: start=0.1, range=0.0001-0.3, kr: start=30, range=0.0001-100; $T_{2f}$: start=0.04, range=0.01-0.2.}

%Additionally it is dependant on $TR$, $T_{RF}$ and $\alpha$, which are however fixed by the acquisition.

%\begin{figure}
%	\centering
%	\makebox[\linewidth][c]{\input{figures/plot22.tex}}
%	\caption[Comparison of qMT bSSFP Signal Equations]{Original (red) and refined (blue) qMT bSSFP signal equation, next to the numerically simulated data (black dots), in a standard acquisition scheme of varied flip angles (left) and pulse durations (right). The plot is exemplary for white matter (parameters in Table \ref{tab:1}) and constant values are $\alpha=35^{\circ}$ and $T_{RF}=\SI{0.3}{ms}$.}
%	\label{fig:simcomp}
%\end{figure}

\section*{Results}

\subsection*{Numerical Studies}

\begin{figure}[hp]
	\centering
	\includegraphics[width=0.83\linewidth]{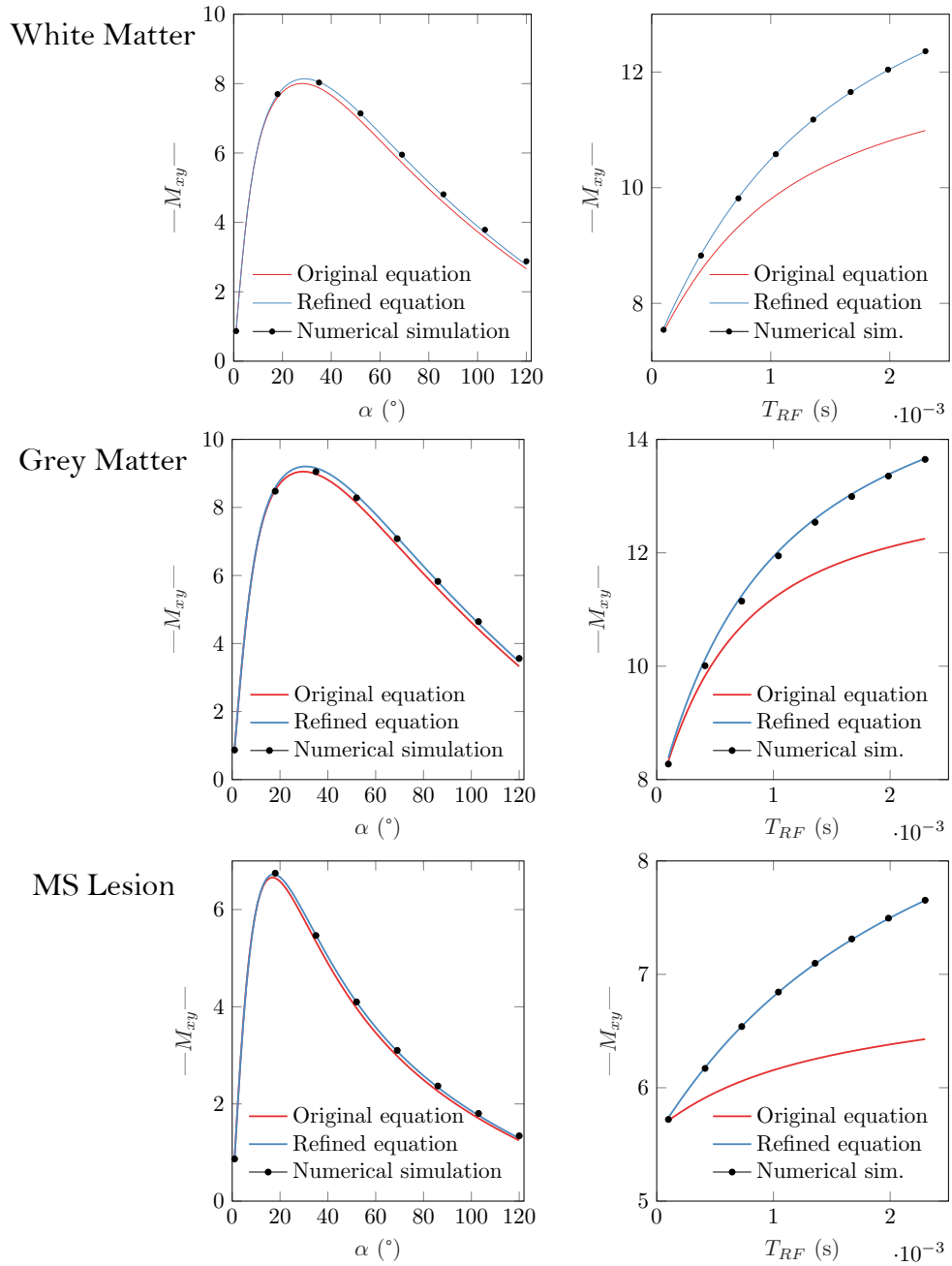}
	\caption[Comparison of qMT bSSFP Signal Equations]{\added{Original (red) and refined (blue) qMT bSSFP signal equation, next to the numerically simulated data (black dots), in a standard acquisition scheme of varied flip angles (left) and pulse durations (right). The plot used the parameters in Table \ref{tab:1} and constant values are $\alpha=35^{\circ}$ and $T_{RF}=\SI{0.3}{ms}$.}}
	\label{fig:simcomp}
\end{figure}

\noindent
Figure \ref{fig:simcomp} shows the original (red) and refined (blue) qMT bSSFP signal equations, along with the numerically simulated data (black dots) for a standard acquisition scheme \added[remark={R1.9}]{for typical brain tissue parameters (Table \ref{tab:1}).}
%While Figure \ref{fig:simcomp} is exemplary for white matter, the corresponding figures for various brain tissue parameters can be found in Appendix \ref{sec:morePlots}.  
The simulation has been performed for a standard acquisition scheme of varied flip angles (left) and pulse durations (right). Taking the numerical simulation as the mathematical ground truth, Figure \ref{fig:simcomp} shows a bias in the original signal equation of up to \SI{11.1}{\percent} at $T_{RF}=\SI{2.3}{ms}$. The maximum bias of both original and refined signal equations for the different tissue parameters are summarised in Table \ref{tab:2}. While the original model is affected by a maximal bias of up to \SI{20.3}{\percent} (for an MS lesion), the refined signal equation describes the numerically simulated data with a bias $<\SI{1}{\percent}$ in all three tissues.

% Discussion: As described in section \ref{sec:derivation}, this bias can be explained by the overestimation of transverse relaxation during excitation in the original signal equation. 

%- plot showing difference in equations for T2/T1 and alpha or Trf/TR and alpha

\subsection*{In-Vivo Studies}

\begin{figure}[hp]
	\centering
	\includegraphics[width=0.88\linewidth]{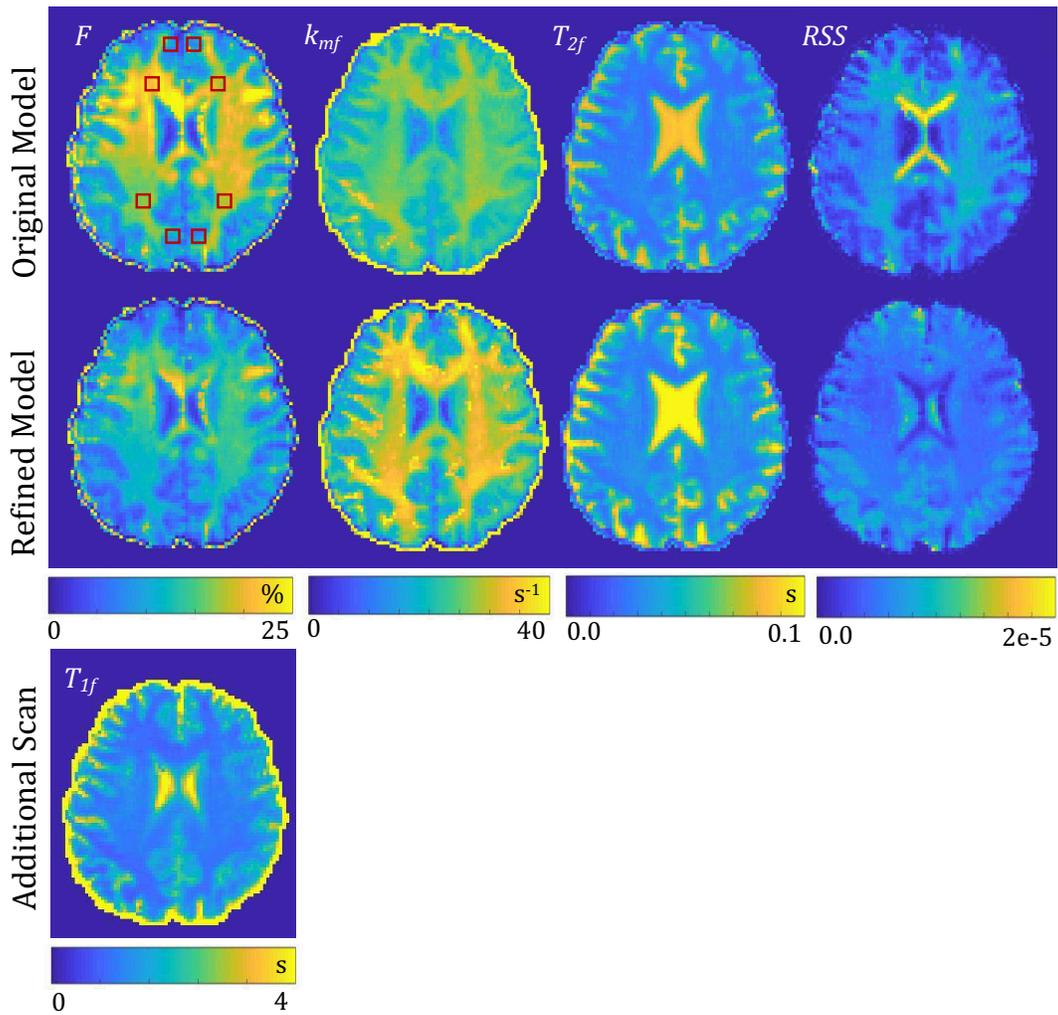}
	\caption[Brain qMT Parameter Maps]{\added{QMT parameter maps of a healthy brain, as analysed by the original (top row) and the refined (bottom row) model. In addition, the residual sum of squares (RSS) of the fit and $T_{1f}$ maps are displayed. The qMT parameters, fitted for each voxel, are as follows: pool-size-ratio $F$, exchange rate $k_{mf}$ and relaxation time of the free pool $T_{2f}$. Red squares mark ROIs.}}
	\label{fig:brain}
\end{figure}

\noindent
The resulting qMT parameter maps of a voxelwise least-squares fit on the human data are shown in Figure \ref{fig:brain}. The analysis was performed using the original and the refined signal equations. The mean values of all fitted qMT parameters within regions of interest (ROIs) in grey and white matter are listed in Table \ref{tab:3}. 

For the clinically relevant pool-size-ratio $F$, a significant difference (p $<$ .001) between both models can be observed within ROIs in white and grey matter. Compared to the original model ($F_{\text{org.}}$), \added[remark={R1.7}]{$F$ decreased in the refined model ($F_\text{ref.}$) from \SI{18.4\pm 1.2}{\percent} / \SI{19.4\pm 0.8}{\percent} to \SI{12.7\pm 0.8}{\percent} / \SI{14.1\pm 0.5}{\percent} in frontal / occipital white matter, respectively.} Similarly, \added[remark={R1.7}]{F decreased from \SI{8.6\pm 1.4}{\percent} to \SI{6.5\pm 1.1}{\percent} and \SI{7.9\pm 0.6}{\percent} to \SI{5.9\pm 0.7}{\percent} in frontal and occipital grey matter, respectively.} The difference between the estimates of the refined and original signal equation is statistically larger (p $<$ .001) in white matter compared to grey matter. 

The exchange rate, analysed by the refined model, $k_{mf,\text{ref.}}$ differs from the original model \added[remark={R2.4}]{}$\added{k_{mf,\text{org.}}}$ only in white matter (p $<$ .001); no statistically significant difference has been found in grey matter (p $=$ .25 and p $=$ .49 in frontal and occipital grey matter, respectively). The refined model results in statistically lower transversal relaxation rates in the free pool $T_{2f}$ (p $<$ .001), with differences ranging from \SI{9}{\percent} - \SI{13}{\percent} in all four ROIs.

The mean of the \added[remark=R1.10]{residual sum of squares (RSS)} over all voxels is \SI{7}{\percent} lower in the refined model compared to the original one.

%- show an image reconstruction fit
%- show the reconstructed images of $F$, $k_{mf}$, $T_{2,f}$ (and additional $T_{1,f}$ in the appendix)
%- discuss (no, don't discuss but include :D) resnorm and potentially include fit
%- include ROI Table of different qMT parameters in areas of brain

\subsection*{Finite Pulse Duration Correction for Sinc-Shape}\label{sec:pusha}

\begin{figure}[hp]
	\centering
%	\makebox[\linewidth][c]{\input{figures/plot19.tex}}
	\includegraphics[width=0.825\linewidth]{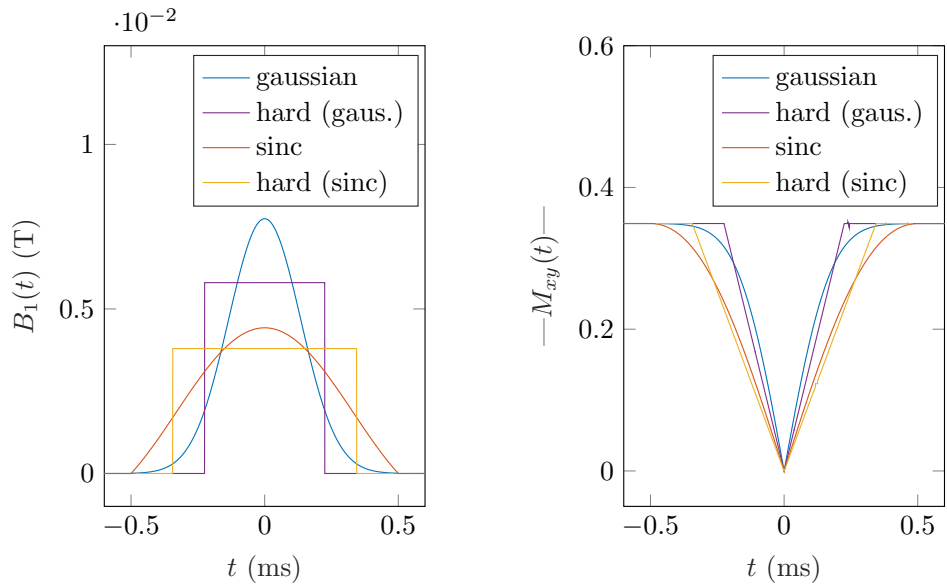}
	\caption[Illustration of Hard Pulse Equivalent]{Illustration of hard pulse equivalent for Gaussian and sinc pulse shapes. The RF pulse (left) and the corresponding transverse magnetization trajectory (right) are plotted for both pulse shapes and their hard pulse equivalent. Both pulse shapes are plotted for $\alpha=40^{\circ}$, $T_{RF}=\SI{1}{ms}$. To allow for a clear distinction, $TBW=2$ for sinc pulse and $TBW=2.6$ for Gaussian pulse.}
	\label{fig:pulse}
\end{figure}

\noindent
Figure \ref{fig:pulse} (left) shows a Gaussian and a sinc pulse of similar flip angle together with their respective hard pulse equivalents. The contributing magnetisations in Figure \ref{fig:pulse} (right) show, that the areas enclosed under the curves are equal for each pulse and its respective hard pulse equivalent. This illustrates the definition of the hard pulse equivalent (Equation \ref{equ:TRFE}). 

Exemplary values of $T_{RFE}$ for the different pulse shapes are listed in Table \ref{tab:Trfe}, showing significant differences for the same $TBW$. The hard pulse equivalent duration approaches zero for $TBW \in \{4,8,12,...\}$ in the sinc pulse. This implies that the correction becomes unnecessary in this case (i.e. $\tilde{R}_2 = R_2$), as the correction term for finite pulse durations directly correlates with the hard pulse equivalent duration $\Delta R_{2f} \propto T_{RFE}$ (Equation \ref{equ:cor}).

The derived relation (Equation \ref{equ:Trefall}) allows correction for the $T_2$-bias in the bSSFP signal equation (both standard and qMT specific) when using sinc pulse shapes. The bias, induced by the overestimation of transversal relaxation during excitation, can be corrected by substituting $R_2\rightarrow \tilde{R}_2$ in the original signal equation. In the case of a sinc pulse shape, the correction factor is as follows
%\begin{equation}
%\tilde{R}_2=\bigg(1-\frac{4 \zeta T_{RF}}{TBW \pi TR} \frac{1- \cos \big(\frac{\pi TBW}{2} \big)}{\text{Si}\big( \frac{\pi TBW}{2} \big)}\bigg)R_2
%\end{equation}
\begin{equation}
\tilde{R}_2=R_{2f}-R_{2f}\frac{4 \zeta T_{RF}}{TBW \pi TR} \frac{1- \cos \big(\frac{\pi TBW}{2} \big)}{\text{Si}\big( \frac{\pi TBW}{2} \big)}
\end{equation}
where 
\begin{equation}
\zeta \approx 0.68-0.125 \bigg(1+ \frac{4 T_{RF}}{TBW \pi TR} \frac{1- \cos \big(\frac{\pi TBW}{2} \big)}{\text{Si}\big( \frac{\pi TBW}{2} \big)} \bigg)  \frac{R_1}{R_2} 
\end{equation}

\section*{Discussion}

\noindent
The simulations have shown that the original qMT bSSFP signal equation is biased by the assumptions made in its derivation (firstly separation of exchange and relaxation and secondly instantaneous rotation of the RF pulse). This bias has been seen to be tissue dependant, amounting to deviations of up to \SI{7}{\percent} and \SI{11}{\percent} in white and grey matter of healthy brain tissue and exceeding \SI{20}{\percent} in an MS lesion. The tissue dependence is expected, as the bias linearly depends on the relaxation time ratio $\frac{T_2}{T_1}$ (Equation \ref{equ:cor}), which varies amongst different tissue types. 
%Though the dependence appears in qMT, it is not a result of it and has also been observed in a previous non-qMT study \cite{bieriSSFPSignalFinite2009}.
Further, the bias has been shown to increase at higher pulse durations $T_{RF}$. This reflects the fact that while the assumption of an instantaneous rotation by the RF pulse might be sufficient for short pulse durations, it is increasingly violated at longer $T_{RF}$. In qMT bSSFP, this bias is particularly strong, as the acquisition involves long $T_{RF}$ relative to $TR$ \cite{gloorQuantitativeMagnetizationTransfer2008}. The bias is passed on to the qMT parameters, as they are determined by fitting the signal equation to the acquired data. 

To address this, the suggested refined signal equation for qMT bSSFP (Equation \ref{equ:Mss}) has been derived, accounting for the assumptions made in the original model, and describes the simulated data with minimal bias ($<$\SI{1}{\percent}). 

The comparison of original and refined signal equations \textit{in-vivo} shows significant differences between the resulting qMT parameters (24-\SI{31}{\percent} for pool-size-ratio, 0-\SI{21}{\percent} for exchange rate and 9-\SI{13}{\percent} for transversal relaxation time). In agreement with the simulation results, the difference between qMT parameters, determined by the original and refined model, is significantly greater in white matter compared to grey matter in \textit{in-vivo} brain tissue data. This is in agreement with the theoretically predicted $\frac{T_2}{T_1}$ dependency of the bias (Equation \ref{equ:cor}).

%The acquired qMT parameters have been compared to the ones found in previous studies using different acquisition techniques. 

In previous studies of different qMT modalities, pool-size-ratios in the range of 10-\SI{16}{\percent} and 3-\SI{8}{\percent} have been reported in white and grey matter structures, respectively \cite{sledQuantitativeImagingMagnetization2001, dortchQuantitativeMagnetizationTransfer2011,dortchOptimizationSelectiveInversion2018,sledRegionalVariationsNormal2004, yarnykhCrossrelaxationImagingReveals2004}. The pool-size-ratios, determined by the original model in this work, exceed the previously reported range in white matter (\SI{18.4\pm1.2}{\percent}, \SI{19.3\pm0.8}{\percent}) and approach the upper limit in grey matter (\SI{8.6\pm1.2}{\percent}, \SI{7.9\pm0.6}{\percent}). In contrast, the refined model estimates of the pool-size-ratio are in good agreement with the findings in other studies in both white matter structures (\SI{12.7\pm0.8}{\percent}, \SI{14.1\pm0.5}{\percent}) and grey matter structures (\SI{6.5\pm1.1}{\percent}, \SI{5.9\pm0.7}{\percent}). This indicates that the refined model outperforms the original one not only in simulation, but also \textit{in-vivo}. This conclusion is further supported by the significantly lower residual sum of squares (RSS) found in the fits of the refined model compared to the original one. The wide range of previously reported exchange rates in different qMT methodologies 10-\SI{40}{s^{-1}} includes the results in both original and refined signal equations in this work.

%Due to low sensitivity of qMT methodologies to the exchange rate, its values between 10-\SI{40}{s^{-1}} have been reported. Therefore, the results of both original and refined signal equations in this work are within the range of previously reported exchange rates.

% Less agreement in the literature can be found in the exchange rate, with reported values between 10-\SI{40}{s^{-1}}. Due to the low sensitivity of qMT methodologies to the exchange rate, the results of both original and refined signal equations in this work are within the range of previously reported exchange rates. 

The pool-size-ratios determined by qMT bSSFP in \cite{gloorQuantitativeMagnetizationTransfer2008} are 13-\SI{16}{\percent} and 6-\SI{7}{\percent} for white and grey matter structures, respectively. Although they fall at the upper end of previous findings, they are lower than the values found with the original model in this work. The reason for reduced biases in the original findings \cite{gloorQuantitativeMagnetizationTransfer2008} can be explained by means of the pulse shape analysis, established in Section \ref{sec:pusha}. While the parameters of the original findings have been acquired with a $TBW=2.7$, in this work a $TBW=2$ has been used. Their respective hard pulse equivalent durations, which correlate with the bias $\Delta R_2 \propto T_{RFE}$, differ by \SI{42}{\percent} $\Big(\frac{T_{RFE}(TBW=2.0)}{T_{RFE}(TBW=2.7)}=0.58\Big)$. This implies a reduction of the bias in the original acquisition scheme for a $TBW=2.7$ and explains why bias is reduced in the original publication \cite{gloorQuantitativeMagnetizationTransfer2008}. While the bias is only reduced and not removed, much higher biases are expected for a $TBW \le 2.5$. Alternatively, the refined signal equation allows for a general solution with accurate parameter estimation for a wide range of $TBW$. 

Additionally, the derived Equation \ref{equ:Trefall} predicts that the bias oscillates for varying $TBW$ and even approaches zero for a $TBW \in \{4,8,12,...\}$. The physical explanation to the oscillation lies in the sinc-shape specific side lobes. %For instance, consider a sinc-shape pulse with a $TBW=4$. Next to the positive main lobe, there are two negative side lobes. 
These side lobes cause a temporary increased deflection of magnetisation from the equilibrium alignment, for which transverse relaxation is underestimated. The underestimation, induced by the negative side lobes, counterbalances the overestimation, resulting from the main lobe. Therefore, the bias in quantitative bSSFP methods (qMT and non-qMT) can be removed by choosing an appropriate time-bandwidth product without using the correction given by Equation \ref{equ:cor}. This might be be useful for applications, where the correction is inaccurate due to strong magnetic field inhomogeneities, such as is the case at high magnetic field strengths.

\added[remark={R1.8}]{Recent work by Wood et al. \cite{wood2020magnetization} has demonstrated that the PLANET method \cite{shcherbakova2018planet, shcherbakova2019accuracy, jones2004squashing} for phase-cycled bSSFP can be applied to qMT at higher field strengths to derive qMT parameter estimates free from banding artefacts. However, Wood et al. \cite{wood2020magnetization} utilised the signal model from Gloor et al. \cite{gloorQuantitativeMagnetizationTransfer2008}, which translated into increased errors in their parameter estimation, particularly in white matter. We hypothesize that combining the method from Wood et al. \cite{wood2020magnetization} with our methods here would provide increased accuracy, leading to a method which can produce qMT parameter estimates quickly over all clinical field strengths. However, this is beyond the scope of this paper, and is left for future work.}

\section*{Conclusion}

\noindent
A new signal equation for qMT bSSFP was derived, which incorporates both finite pulse effects and simultaneous magnetization exchange and relaxation. Numerical simulations of the Bloch-McConnell equations showed that the original signal equation is significantly biased in typical brain tissue structures (by 7-20 \%). By contrast, the new signal equation outperforms the original one with minimal bias ($<$ 1 \%). The practicality of the new signal equation was demonstrated using \textit{in-vivo} data and it is shown that the bias of the original signal equation strongly affects the acquired qMT parameters in human brain structures, with differences in the clinically relevant pool-size-ratio of up to 31 \%. Particularly high biases of the original signal equation are expected in an MS lesion within diseased brain tissue (due to a low $\frac{T_{2f}}{T_{1f}}$-ratio), demanding a more accurate model for clinical applications. Therefore, the refined signal equation is recommended for accurate qMT parameter estimation in healthy and diseased brain tissue, especially when using a $TBW \le 2.5$.

\section*{Acknowledgements}

\noindent
The Wellcome Centre for Integrative Neuroimaging is supported by core funding from the Wellcome Trust (203139/Z/16/Z). We also thank the Dunhill Medical Trust and the NIHR Oxford Biomedical Research Centre for support (PJ). AKS acknowledges support from the Whitaker International Program and St. Hilda’s College at the University of Oxford. FMB acknowledges support from the Cusanuswerk Scholarship. 

\appendix
\setcounter{equation}{0}
\renewcommand\theequation{A.\arabic{equation}}

\section*{Appendix: Finite Pulse Duration Correction for Sinc-Shape}\label{sec:TRFE}
%\subsection{Finite RF Pulse Duration Correction for Sinc-Shape}\label{sec:TRFE}

\noindent
As shown by Bieri \cite{bieriSSFPSignalFinite2009,bieriAnalyticalDescriptionBalanced2011}, the overestimation of transverse relaxation in bSSFP can be corrected by means of the substitution of $R_2 \rightarrow \tilde{R}_2$  according to Equation \ref{equ:cor}. This correction has been derived under the assumption of excitation by a constant RF magnetic field (i.e. a hard pulse). By means of the hard pulse equivalent duration $T_{RFE}$, the correction can be transferred to different pulse shapes
\begin{equation}\label{equ:hpe}
T_{RFE}  \braket{B} = \int B_1(t)dt
\end{equation}
where $<B>$ is the mean $B_1$ amplitude. This relation has previously been solved for a Gaussian pulse shape, resulting in $T_{RFE}=1.2 \: \frac{T_{RF}}{TR}$ \cite{bieriSSFPSignalFinite2009}. 
%However, Gaussian pulse shapes are not commonly used in qMT bSSFP, where a sinc pulse shape is suggested \cite{gloorQuantitativeMagnetizationTransfer2008, garciaFastHighresolutionBrain2012, gloorIntrascannerInterscannerVariability2011}. %Thus, in order to use the correction in qMT bSSFP, Ansatz \ref{equ:TRFE} has been solved for a sinc pulse shape. 
In this section, Ansatz \ref{equ:hpe} is solved for a sinc pulse shape. 
%\\
%%\noindent
%%\textbf{Note:} While the previous section has been based on qMT bSSFP, the following derivation is made for bSSFP in general.
%\noindent
%\textit{Derivation}

%\noindent
Consider a sinc pulse of form
\begin{equation}\label{equ:sinc}
B_1(t)=	\begin{cases} 
A \, \text{sinc}\big(\frac{\pi t}{t_0}\big)  &\;\; \text{for} -N_L t_0 \le t \le N_R t_0 \\
0 &\;\; \text{elsewhere}
\end{cases}
\end{equation}
where $t_0$ is the half-width of the central lobe, $A$ is the amplitude of the pulse and $t$ is the time. $N_L$ and $N_R$ are the numbers of zero-crossings of the sinc pulse to the left and right of the central peak, respectively (if symmetric: $N_L=N_R$). The full width at half maximum (FWHM) of a sinc pulse can be approximated by $\text{FWHM} = \Delta f \approx \frac{1}{t_0}$  \cite{bernsteinHandbookMRIPulse2004} and the time bandwidth product is defined as 
\begin{equation}\label{equ:TBW}
TBW= N_L + N_R = N =T_{RF} \Delta f \approx \frac{T_{RF}}{t_0}
\end{equation}
with the number of total zero-crossings $N$. 

As shown by Bieri \cite{bieriSSFPSignalFinite2009}, Ansatz \ref{equ:hpe} leads to a relation between $T_{RFE}$ and magnetisation
\begin{equation}\label{equ:TRFE}
\langle M_{xy} \rangle ^+ = \bigg(1- \frac{T_{RFE}}{2T_{RF}}\bigg) M_{xy}^-
\end{equation}
where $M_{xy}^-$ is the magnetisation before excitation and $\langle M_{xy} \rangle ^+$ is the time average magnetisation during excitation. By calculating the pulse shape dependent $\langle M_{xy} \rangle ^+$, Equation \ref{equ:TRFE} can be used to find the hard pulse equivalent duration $T_{RFE}$. For sufficiently small flip angles, the magnetisation can be approximated to
\begin{equation}
M_{xy}(t) \approx \frac{\bigl| \alpha (t)- \alpha/2 \bigl|}{\alpha / 2} \cdot M_{xy}^- = \biggl| \gamma \int_{-T_{RF}/2}^{t} \frac{B_1(t')}{\alpha / 2} dt' - 1 \biggl| \cdot M_{xy}^-
\end{equation}
where the the relations of flip angle $\alpha(t)= \int_{-T_{RF}/2}^{t} \omega_1(t') dt'$ and $\omega_1(t)=\gamma B_1(t)$, have been used.

Using the symmetry property of the sinc function $\text{sinc}(x)=\text{sinc}(-x)$, the flip angle can be expressed as
\begin{equation}
\alpha=\gamma \int_{-T_{RF}/2}^{T_{RF}/2} B_1(t)dt=2 \gamma \int_{0}^{T_{RF}/2} B_1(t)dt=2 \gamma \frac{A t_0}{\pi} \int_{0}^{\frac{T_{RF} \pi }{2 t_0}} \text{sinc}\big( \theta \big) d\theta
\end{equation}
where the substitution $\frac{\pi t}{t_0}= \theta $ has been used. The integral can be solved using the sine integral definition
\begin{equation}\label{equ:sineint}
\text{Si}\big( x \big) = \int_{0}^{x} \frac{\sin(\theta)}{\theta}d\theta
\end{equation}
leading to a relation between flip angle, pulse duration and half-bandwidth
\begin{equation}
\alpha = 2 \gamma \frac{A t_0}{\pi} \: \text{Si}\Big( \frac{T_{RF} \pi}{2 t_0}\Big)
\end{equation}
Calculating the time average magnetisation during excitation, and exploiting the symmetry around $t=0$, results in
\begin{equation}\label{equ:magg}
\langle M_{xy} \rangle ^+ = \frac{1}{T_{RF}} \int_{-T_{RF}/2}^{T_{RF}/2} M_{xy}(t)dt = \frac{2}{T_{RF}} \int_{0}^{T_{RF}/2} \biggl| \gamma \int_{-T_{RF}/2}^{t} \frac{B_1(t')}{\alpha / 2} dt' - 1 \biggl| \cdot M_{xy}^- dt
\end{equation}
The integral of $B_1(t)$ can be split $\forall \: t \in [0,T_{RF}/2]$ into
\begin{equation}
\int_{-T_{RF}/2}^{t} B_1(t')dt'= \int_{-T_{RF}/2}^{0} B_1(t')dt'+\int_{0}^{t} B_1(t')dt'=\frac{\alpha}{2\gamma}+\int_{0}^{t} B_1(t')dt'
\end{equation}
where the definition of flip angle $\alpha$ and the symmetric nature of the sinc function have been used.
This simplifies Equation \ref{equ:magg} to 
\begin{equation}
\begin{aligned}
\langle M_{xy} \rangle ^+ 
& = \frac{2}{T_{RF}} \int_{0}^{T_{RF}/2} \biggl| \frac{2\gamma}{\alpha} \bigg( \frac{\alpha}{2\gamma}+\int_{0}^{t} B_1(t')dt' \bigg) - 1 \biggl| \cdot M_{xy}^- dt\\
& = \frac{4 \gamma A}{T_{RF} \alpha} \int_{0}^{T_{RF}/2} \biggl| \int_{0}^{t} \text{sinc} \bigg(\frac{\pi t'}{t_0} \bigg) dt' \biggl| \cdot M_{xy}^- dt
\end{aligned}
\end{equation}
Further substitutions and the use of the integral in Equation \ref{equ:sineint} leads to
%\begin{equation}
%	\begin{aligned}
%	\langle M_{xy} \rangle ^+ 
%	& = \frac{4 \gamma A}{T_{RF} \alpha} \frac{t_0}{\pi} \int_{0}^{T_{RF}/2} \biggl| \text{Si} \bigg(\frac{\pi t}{t_0} \bigg) \biggl| \cdot M_{xy}^- dt\\
%	& = \frac{4 \gamma A}{T_{RF} \alpha} \bigg(\frac{t_0}{\pi}\bigg)^2 \int_{0}^{\frac{T_{RF} \pi}{2 t_0}} \biggl| \text{Si} \big(u \big) \biggl| \cdot M_{xy}^- du
%	\end{aligned}
%\end{equation}
\begin{equation}
\langle M_{xy} \rangle ^+ = \frac{4 \gamma A}{T_{RF} \alpha} \frac{t_0}{\pi} \int_{0}^{T_{RF}/2} \biggl| \text{Si} \bigg(\frac{\pi t}{t_0} \bigg) \biggl| \cdot M_{xy}^- dt = \frac{4 \gamma A}{T_{RF} \alpha} \bigg(\frac{t_0}{\pi}\bigg)^2 \int_{0}^{\frac{T_{RF} \pi}{2 t_0}} \biggl| \text{Si} \big(u \big) \biggl| \cdot M_{xy}^- du
\end{equation}
The integral of the sine integral $\text{Si}\big(t\big)$ is found by means of partial integration
\begin{equation}
\int_{0}^{T}\text{Si}\big(t\big)dt= \int_{0}^{T}1 \cdot \text{Si}\big(t\big)dt\stackrel{\text{PI}}{=}\Big[t \:  \text{Si}\big(t\big)\Big]_0^T - \int_{0}^{T} \sin \big(t \big) dt = T \: \text{Si}\big(T\big) + \cos \big( T \big) - 1 
\end{equation}
to yield 
\begin{equation}\label{equ:maggg}
\langle M_{xy} \rangle ^+ = \bigg[1+\frac{2}{TBW \pi} \frac{1}{\text{Si}\big(\frac{\pi TBW}{2}\big)} \bigg( \cos \Big(\frac{\pi TBW}{2} \Big) - 1 \bigg) \bigg] M_{xy}^-
\end{equation}
where relation \ref{equ:TBW} has been used. 

Inserting the derived expression for the time average magnetisation \ref{equ:maggg} into Ansatz \ref{equ:TRFE} finally leads to the relation  
\begin{equation}\label{equ:Trefsinc}
T_{RFE}= \frac{4 T_{RF}}{TBW \pi} \frac{1- \cos \big(\frac{\pi TBW}{2} \big)}{\text{Si}\big( \frac{\pi TBW}{2} \big)}
\end{equation}
%\qed

This relation allows for correction of overestimation of transversal relaxation in bSSFP, when using a sinc pulse for excitation. Analogous to the correction for Gaussian and hard pulse shapes, a substitution $R_2 \rightarrow \tilde{R}_2(T_{RFE})$ corrects for the bias according to Equation \ref{equ:cor}.

%\section{Appendix: Bias in qMT bSSFP Signal Equation}\label{sec:morePlots}

%\begin{figure}[H]
%	\centering
%	\makebox[\linewidth][c]{\input{figures/plot23.tex}}
%	\caption[QMT bSSFP signal equation in MS lesion]{Original (red) and the refined (blue) qMT bSSFP signal equation, next to the numerically simulated data, in a standard acquisition scheme of varied flip angles (left) and pulse durations (right). The plot is exemplary for MS lesion (Table \ref{tab:1}).}
%	\label{fig:appendixMS}
%	\vspace*{\floatsep}% https://tex.stackexchange.com/q/26521/5764
%	
%	\makebox[\linewidth][c]{\input{figures/plot20.tex}}
%	\caption[QMT bSSFP signal equation in grey matter]{Original (red) and the refined (blue) qMT bSSFP signal equation, next to the numerically simulated data, in a standard acquisition scheme of varied flip angles (left) and pulse durations (right). The plot is exemplary for grey matter (Table \ref{tab:1}).}
%	\label{fig:appendixGM}
%\end{figure}

% include two references as suggested by reviewer R1.2
\nocite{crooijmans2011influence, crooijmans2011finiteb}

\bibliography{main.bib}\added[remark={R1.2}]{}

\clearpage

\listoffigures

\listoftables

% all figures here

%\begin{figure}[hp]
%	\centering
%%	\makebox[\linewidth][c]{\input{figures/plot22.tex}}
%	\includegraphics[width=0.83\linewidth]{fig1.eps}
%	\caption[Comparison of qMT bSSFP Signal Equations]{Original (red) and refined (blue) qMT bSSFP signal equation, next to the numerically simulated data (black dots), in a standard acquisition scheme of varied flip angles (left) and pulse durations (right). The plot is exemplary for white matter (parameters in Table \ref{tab:1}) and constant values are $\alpha=35^{\circ}$ and $T_{RF}=\SI{0.3}{ms}$.}
%	\label{fig:simcomp}
%\end{figure}

%\begin{figure}[hp]
%	\centering
%	\includegraphics[width=0.98\linewidth]{brain_cabana_org_ref_version1.pdf}
%	\caption[Brain qMT Parameter Maps]{QMT parameter maps of a healthy brain, as analysed by the original (top row) and the refined (bottom row) model. The qMT parameters, fitted for each voxel, are as follows: pool-size-ratio $F$, exchange rate $k_{mf}$ and relaxation time of the free pool $T_{2f}$. Red squares mark ROIs.}
%	\label{fig:brain}
%\end{figure}

%\begin{figure}[hp]
%	\centering
%	\includegraphics[width=0.75\linewidth]{fig2.eps}
%	\caption[Brain qMT Parameter Maps]{QMT parameter maps of a healthy brain, as analysed by the original (top row) and the refined (bottom row) model. The qMT parameters, fitted for each voxel, are as follows: pool-size-ratio $F$, exchange rate $k_{mf}$ and relaxation time of the free pool $T_{2f}$. Red squares mark ROIs.}
%	\label{fig:brain}
%\end{figure}

\renewcommand\thefigure{\thesection.\arabic{figure}} 
\setcounter{figure}{0}

%\begin{figure}[hp]
%	\centering
%%	\makebox[\linewidth][c]{\input{figures/plot23.tex}}
%	\includegraphics[width=0.825\linewidth]{fig4.eps}
%	\caption[QMT bSSFP signal equation in MS lesion]{Original (red) and the refined (blue) qMT bSSFP signal equation, next to the numerically simulated data, in a standard acquisition scheme of varied flip angles (left) and pulse durations (right). The plot is exemplary for MS lesion (Table \ref{tab:1}).}
%	\label{fig:appendixMS}
%	\vspace*{\floatsep}% https://tex.stackexchange.com/q/26521/5764
%	
%%	\makebox[\linewidth][c]{\input{figures/plot20.tex}}
%	\includegraphics[width=0.825\linewidth]{fig5.eps}
%	\caption[QMT bSSFP signal equation in grey matter]{Original (red) and the refined (blue) qMT bSSFP signal equation, next to the numerically simulated data, in a standard acquisition scheme of varied flip angles (left) and pulse durations (right). The plot is exemplary for grey matter (Table \ref{tab:1}).}
%	\label{fig:appendixGM}
%\end{figure}

% all tables here
\FloatBarrier

\begin{table}[hp]
	\centering
	\caption[Typical qMT Tissue Parmeters]{Typical qMT tissue parameters for different areas of the brain, taken from \cite{dortchOptimizationSelectiveInversion2018,garciaFastHighresolutionBrain2012}. $T_{2f}$: longitudinal relaxation time of the free pool.}\label{tab:1}
	\setlength{\tabcolsep}{15pt}
	\begin{tabular}{SSSSS}	
		\toprule
		{Tissue} & {$F$ (\si{\percent})}&{$k_{mf}$ (\si{s}$^{-1}$)}&{$R_{1f}$ (\si{s}$^{-1}$)}&{$T_{2f}$ (\si{ms})}\\
		\midrule
		{White matter} & {11}&{\SI{10}{}}&{\SI{0.9}{}}&{\SI{42}{}}\\
		{Grey matter} & {6}&{\SI{18}{}}&{\SI{0.8}{}}&{\SI{74}{}}\\
		{MS lesion} & {3}&{\SI{8}{}}&{\SI{0.5}{}}&{\SI{43}{}}\\
		\bottomrule
	\end{tabular}
\end{table}

\begin{table}[hp]
	\centering
	\caption[Maximal Percentage Deviation of Analytical Signal Equations from the Numerical Simulation]{Maximal percentage deviation of analytical signal equations from the numerical simulation $\Delta M_{y,\text{max.}}=\max \big[ (M_{y,\text{simulation}}-\Delta M_{y,\text{analytical}})/M_{y,\text{simulation}}\big]$ in the standard protocol of varied flip angles and pulse durations. The maximum has been calculated for $T_{RF}$ ranging from \SI{0.2}{ms} to \SI{2.3}{ms}, $\alpha$ ranging from $5^{\circ}$ to $40^{\circ}$, fixed $t_d=TR-T_{RF}=\SI{2}{ms}$, $TBW=2$ and the qMT parameters of Table \ref{tab:1}.}\label{tab:2}
	\setlength{\tabcolsep}{17pt}
	\begin{tabular}{SSS}	
		\toprule
		{Tissue} & {Original bias $\Delta M_{y, \text{max.}}$}&{Refined bias $\Delta M_{y, \text{max.}}$}\\
		\midrule
		{White matter} & {\SI{11.1}{\percent}}&{\SI{0.7}{\percent}}\\
		{Grey matter} & {\SI{7.4}{\percent}}&{\SI{0.3}{\percent}}\\
		{MS lesion} & {\SI{20.3}{\percent}}&{\SI{0.4}{\percent}}\\
		\bottomrule
	\end{tabular}
\end{table}

\begin{table}[hp]
	\centering
	\caption[Fitted qMT Parameters Determined by the Original and the Refined Model]{Fitted qMT parameters within healthy brain structures, as determined by the original and the refined model. Shown are the mean $\pm$ standard deviation (SD) values across the ROIs. WM: white matter, GM: grey matter.}\label{tab:3}
	\setlength{\tabcolsep}{2.0pt}
	\begin{tabular}{SSSSSSS}	
		\toprule
		{Tissue} & {{$F_{\text{org.}}$ (\si{\percent})}}&{{$F_{\text{ref.}}$ (\si{\percent})}} & {{$k_{mf,\text{org.}}$ (s$^{-1}$)}} & {{$k_{mf,\text{ref.}}$ (s$^{-1}$)}} & {{$T_{2f,\text{org.}}$ (ms)}} & {{$T_{2f,\text{ref.}}$ (ms)}}\\
		\midrule
		{Frontal WM} & {\SI{18.4 \pm 1.2}{}}&{\SI{12.7 \pm 0.8}{}}& {\SI{28.2 \pm 0.6}{}}&{\SI{35.6 \pm 1.3}{}}& {\SI{31 \pm 5}{}}&{\SI{27 \pm 4}{}}\\
		{Frontal GM} & {\SI{8.6 \pm 1.4}{}}&{\SI{6.5 \pm 1.1}{}}& {\SI{22.2 \pm 2.8}{}}&{\SI{22.9 \pm 4.9}{}}& {\SI{62 \pm 11}{}}&{\SI{54 \pm 9}{}}\\
		{Occipital WM} & {\SI{19.4 \pm 0.8}{}}&{\SI{14.1 \pm 0.5}{}}& {\SI{27.2 \pm 0.8}{}}&{\SI{33.1 \pm 1.3}{}}& {\SI{32 \pm 2}{}}&{\SI{28 \pm 2}{}}\\
		{Occipital GM} & {\SI{7.9 \pm 0.6}{}}&{\SI{5.9 \pm 0.7}{}}& {\SI{24.0 \pm 3.7}{}}&{\SI{25.2 \pm 8.6}{}}& {\SI{32 \pm 6}{}}&{\SI{29 \pm 5}{}}\\
		\bottomrule
	\end{tabular}
\end{table}

\begin{table}[hp]
	\centering
	\caption[Exemplary Hard Pulse Equivalent Duration of Gaussian, Sinc and Hard Pulse Shapes]{Exemplary hard pulse equivalent duration $T_{RFE}$ of Gaussian, sinc and hard pulse shapes for different $TBW$, resulting from Equation \ref{equ:Trefall}.}\label{tab:Trfe}
	\setlength{\tabcolsep}{10pt}
	\begin{tabular}{SSSS}	
		\toprule
		{\textbf{$TBW$}} & {Sinc pulse $T_{RFE}$}&{Gaussian pulse $T_{RFE}$}&{Hard pulse $T_{RFE}$}\\
		\midrule
		%		{1} & {0.93 $T_{RF}$}&{1.20 $T_{RF}$}&{1.00 $T_{RF}$}\\
		{2} & {0.69 $T_{RF}$}&{0.60 $T_{RF}$}&{1.00 $T_{RF}$}\\
		{3} & {0.26 $T_{RF}$}&{0.40 $T_{RF}$}&{1.00 $T_{RF}$}\\
		{4} & {0}&{0.30 $T_{RF}$}&{1.00 $T_{RF}$}\\
		%		{5} & {0.16 $T_{RF}$}&{0.24 $T_{RF}$}&{1.00 $T_{RF}$}\\
		%		{6} & {0.25 $T_{RF}$}&{0.2 $T_{RF}$}&{1.00 $T_{RF}$}\\
		\bottomrule
	\end{tabular}
\end{table}

\end{document}